\documentclass[preprintnumbers,secnumarabic]{revtex4}
\pdfoutput=1

\usepackage{graphicx}
\usepackage{amsmath,amssymb,mathrsfs}
\usepackage{fancyhdr}
\usepackage{psfrag}
\usepackage{array,bbm}
\usepackage{rotating}
\usepackage{multirow}
\usepackage{tabularx} 

\newcommand*{\trans}{\mathrm{T}}                     
\newcommand*{\unitmatrix}{\mathbbm{1}}

\newcommand*{\tvec}[1]{\ensuremath{\boldsymbol{\mathrm{#1}}}}           
\newcommand*{\tmat}[1]{\underline{#1}}             

\DeclareMathOperator{\im}{Im}

\DeclareMathOperator{\diag}{diag}

\begin{document}

\title{Symmetries and renormalisation in two-Higgs-doublet models}

\preprint{BI-TP 2011/14}

\author{M. Maniatis}
    \email[E-mail: ]{Maniatis@physik.uni-bielefeld.de}
\affiliation{Fakult\"at f\"ur Physik, Universit\"at Bielefeld,
33615 Bielefeld, Germany}
\author{O. Nachtmann}
    \email[E-mail: ]{O.Nachtmann@thphys.uni-heidelberg.de}
\affiliation{
Institut f\"ur Theoretische Physik, Philosophenweg 16, 69120
Heidelberg, Germany
}

\begin{abstract}
We discuss the classification of symmetries and the corresponding
symmetry groups in the
two-Higgs-doublet model~(THDM). 
We give an easily useable method how
to determine the 
symmetry class and corresponding symmetry group of a given THDM Higgs potential.
One of the symmetry classes
corresponds to a Higgs potential with several simultaneous
generalised CP symmetries. 
Extending the CP symmetry of this class
to the Yukawa sector in a straightforward way,
the so-called maximally-CP-symmetric model~(MCPM)
is obtained.
We study the evolution of
the quartic Higgs-potential parameters
under a change of renormalisation point.
Finally we 
compute the so called
oblique parameters $S$, $T$, and $U$, in the MCPM 
and we identify large
regions of viable parameter space with
respect to electroweak precision measurements.
We present the corresponding allowed regions
for the masses of the physical Higgs bosons.
Reasonable ranges for these masses,
up to several hundred GeV,
are obtained which should make
the (extra) Higgs bosons detectable in LHC experiments.
\end{abstract}

\maketitle

%
\section{Introduction}
\label{sec_intro}

In today's particle physics one of the main hunting grounds of
theorists and experimentalists alike are {\em scalars}.
In the Standard Model~(SM) we have as scalar one
Higgs-boson doublet field,
playing an essential role. It is supposed to be responsible
for electroweak symmetry breaking thereby giving mass
to the~$W$ and $Z$ bosons as well as to quarks and leptons.
However, more complicated Higgs sectors are by no means
excluded experimentally.
On the contrary, there are good theoretical reasons for more than 
one Higgs-boson doublet field.
Extended Higgs sectors are, for instance, required in 
supersymmetric models; see for instance~\cite{Fayet:1974fj,Fayet:1974pd,
Inoue:1982pi,Inoue:1983pp,Flores:1982pr,Gunion:1984yn},
and in many models trying to solve the so called
{\em strong CP problem}~\cite{Peccei:1977hh,Peccei:1977ur}.

One simple extension of the SM scalar sector has
two Higgs-boson doublet fields. This two-Higgs-doublet 
model~(THDM) has been studied extensively in the 
literature; see~\cite{Lee:1973iz,Deshpande:1977rw,Georgi:1978xz,Haber:1978jt,
Donoghue:1978cj,Golowich:1978nh,Hall:1981bc,Haber:1993an,Cvetic:1993cy,
Botella:1994cs,Lavoura:1994fv,Lavoura:1994yu,Velhinho:1994vh,Bernreuther:1998rx,
Davidson:2005cw,Ginzburg:2004vp,
Barbieri:2005kf,Nagel:2004sw,Maniatis:2006fs,Nishi:2006tg,
Ivanov:2006yq,Maniatis:2007vn,Ferreira:2009wh,Mahmoudi:2009zx,
Grzadkowski:2010dj} and references therein.
In our group we have, in particular, emphasised the
usefulness of gauge-invariant bilinears for studying
properties of THDMs and we have introduced a special
THDM, the maximally CP-symmetric model~(MCPM) which 
may give some understanding of the family structure and the fermion
mass hierarchies observed in Nature~\cite{Maniatis:2007de}.
Predictions of the MCPM for high-energy
proton--antiproton and proton--proton collisions
were presented in~\cite{Maniatis:2009vp,Maniatis:2009by,Maniatis:2010sb}.

THDM's with additional symmetries were studied in~\cite{Ivanov:2007de,Ma:2009ax,
Ferreira:2010hy}.
A review of the relation between the usual field formalism
and the geometric picture for THDMs working with
field bilinears was given in~\cite{Ferreira:2010yh}.

In the present work we make some remarks concerning symmetries
and the corresponding groups for THDMs.
We discuss the renormalisation procedure in
view of the symmetry constraints
on the potential parameters.
As an explicit example we treat the renormalisation
of the dimension-four couplings in the MCPM.
Finally we calculate the so called
oblique parameters $S$, $T$, $U$~\cite{Peskin:1990zt}
for the MCPM. Comparing with electroweak
precision data we derive restrictions
on the masses of the 
(extra compared to the SM)
Higgs bosons for the MCPM.

%
\section{The bilinear formalism}
\label{sec_bilinear}

We consider models with the particle content
as in the SM but with two Higgs-boson doublets
\begin{equation}
\label{eq0}
\varphi_i(x) =
\begin{pmatrix}
\varphi_i^+(x)\\ \varphi_i^0(x)
\end{pmatrix},
\end{equation}
where $i=1,2$. Both doublets are assigned weak
hypercharge $y=1/2$.
We use the conventions for 
kinematics etc. as
in \cite{Maniatis:2007de}.
The most general gauge invariant and renormalisable 
potential of the THDM may be written in terms of fields as
\cite{Haber:1993an}
\begin{equation}
\label{eqV}
\begin{split}
V (\varphi_1, \varphi_2)=&
\phantom{+}m_{11}^2 (\varphi_1^\dagger \varphi_1) +
m_{22}^2 (\varphi_2^\dagger \varphi_2) 
-m_{12}^2 (\varphi_1^\dagger \varphi_2) -
(m_{12}^2)^* (\varphi_2^\dagger \varphi_1)
\\
&
+\frac{\lambda_1}{2} (\varphi_1^\dagger \varphi_1)^2
+ \frac{\lambda_2}{2} (\varphi_2^\dagger \varphi_2)^2
+ \lambda_3 (\varphi_1^\dagger \varphi_1)(\varphi_2^\dagger \varphi_2)
+ \lambda_4 (\varphi_1^\dagger \varphi_2)(\varphi_2^\dagger \varphi_1)
+ \frac{1}{2} [\lambda_5 (\varphi_1^\dagger \varphi_2)^2 + \lambda_5^*
(\varphi_2^\dagger \varphi_1)^2]
\\
&
+ [\lambda_6 (\varphi_1^\dagger \varphi_2) + \lambda_6^*
(\varphi_2^\dagger \varphi_1)] (\varphi_1^\dagger \varphi_1) 
+ [\lambda_7 (\varphi_1^\dagger
\varphi_2) + \lambda_7^* (\varphi_2^\dagger \varphi_1)] (\varphi_2^\dagger \varphi_2)\;,
\end{split}
\end{equation}
with $m_{11}^2$, $m_{22}^2$, $\lambda_{1,2,3,4}$ real,
$m_{12}^2$, $\lambda_{5,6,7}$ complex.
To study the properties of the Higgs potential,
it is convenient to write it in terms of
field bilinears \cite{Nagel:2004sw,
Maniatis:2006fs,Nishi:2006tg,Ivanov:2006yq}.
In~\cite{Nagel:2004sw, Maniatis:2006fs}
a one-to-one correspondence of bilinear
gauge-invariant expressions with a
Minkowski-type four vector was revealed leading
to a simple
geometric interpretation.
We arrange the fields~$\varphi_i$ 
of~\eqref{eq0} in a $2 \times 2$ matrix
\begin{equation}
\phi(x) =
\begin{pmatrix}
\varphi_1^+(x) & \varphi_1^0(x)\\
\varphi_2^+(x) & \varphi_2^0(x)
\end{pmatrix}
\end{equation}
and define
the hermitian, positive semi definite, 
$2 \times 2$ matrix
\begin{equation}
\label{eq-kmat}
\tmat{K}(x) := \phi(x) \phi^\dagger(x) =
\begin{pmatrix}
  \varphi_1^{\dagger}(x)\varphi_1(x) & \varphi_2^{\dagger}(x)\varphi_1(x) \\
  \varphi_1^{\dagger}(x)\varphi_2(x) & \varphi_2^{\dagger}(x)\varphi_2(x)
\end{pmatrix}\,.
\end{equation}
Its decomposition reads
\begin{equation}\label{2.4}
\tmat{K}(x)
 = \frac{1}{2}\left( K_0(x)\unitmatrix_2
      + \tvec{K}(x)\,\tvec{\sigma} \right)
\end{equation}
with Pauli matrices $\sigma^a\ (a=1,2,3)$. In this way one defines
the real bilinears
\begin{equation}
\label{eq-kdef}
K_0(x) = \varphi_1^{\dagger} \varphi_1 + \varphi_2^{\dagger} \varphi_2, \quad 
K_1(x) = \varphi_1^\dagger \varphi_2 + \varphi_2^\dagger \varphi_1, \quad
K_2(x) = i \varphi_2^\dagger \varphi_1 - i \varphi_1^\dagger \varphi_2, \quad
K_3(x) = \varphi_1^{\dagger} \varphi_1 - \varphi_2^{\dagger} \varphi_2 \,.
\end{equation}
We have 
\begin{equation}
K_0(x) \ge 0 , \qquad
\left( K_0(x) \right)^2 - \left( \tvec{K}(x) \right)^2 \ge 0.
\end{equation}
In terms of these bilinears the general THDM potential~\eqref{eqV} 
can be written in the simple form
\begin{equation}
\label{eqVK}
V (\varphi_1, \varphi_2)=
  \; \xi_0\, K_0(x) + \tvec{\xi}^\trans\, \tvec{K}(x)
  + \eta_{00}\, K_0^2(x) 
  + 2\, K_0(x)\,\tvec{\eta}^\trans\, \tvec{K}(x)
  + \tvec{K}^\trans(x)\, E\, \tvec{K}(x)\,,
\end{equation}
with $\tvec{K}(x)=(K_1(x), K_2(x), K_3(x))^\trans$ and parameters
$\xi_0$, $\eta_{00}$, three-component vectors $\tvec{\xi}$, $\tvec{\eta}$
and the $3\times 3$ matrix $E=E^\trans$.
All parameters in~\eqref{eqVK} are real. 
The translation from the conventional
parameters to
the bilinear parameters 
is 
\begin{gather}
\begin{split}
\label{connect}
\xi_0=\frac{1}{2}
(m_{11}^2+m_{22}^2)\;,
\quad
\tvec{\xi}=\frac{1}{2}
\begin{pmatrix}
- 2 \textrm{Re}(m_{12}^2)\\
2 \textrm{Im}(m_{12}^2)\\
 m_{11}^2-m_{22}^2
\end{pmatrix},
\\
\eta_{00} =
\frac{1}{8}(\lambda_1 + \lambda_2) + \frac{1}{4}\lambda_3\;,
\quad
\tvec{\eta}=\frac{1}{4}
\begin{pmatrix}
\textrm{Re}(\lambda_6+\lambda_7)\\
-\textrm{Im}(\lambda_6+\lambda_7)\\
\frac{1}{2}(\lambda_1 - \lambda_2)
\end{pmatrix},
\\
E = \frac{1}{4}
\begin{pmatrix}
\lambda_4 + \textrm{Re}(\lambda_5) &
-\textrm{Im}(\lambda_5) &
\textrm{Re}(\lambda_6-\lambda_7) \\
-\textrm{Im}(\lambda_5) &
\lambda_4 - \textrm{Re}(\lambda_5) &
-\textrm{Im}(\lambda_6-\lambda_7) \\
\textrm{Re}(\lambda_6-\lambda_7) &
-\textrm{Im}(\lambda_6 -\lambda_7) &
\frac{1}{2}(\lambda_1 + \lambda_2) - \lambda_3
\end{pmatrix}.
\end{split}
\end{gather}
We also define the $4 \times 4$~ matrix of the 
parameters corresponding to the potential terms
quadratic in the bilinears,
\begin{equation}
\label{Kquart}
\tilde{E} =
\begin{pmatrix}
\eta_{00} & \tvec{\eta}^\trans \\
\tvec{\eta} & E
\end{pmatrix} =
\begin{pmatrix}
\eta_{00} & \eta_{01} & \eta_{02} & \eta_{03} \\
\eta_{01} & \eta_{11} & \eta_{12} & \eta_{13} \\
\eta_{02} & \eta_{12} & \eta_{22} & \eta_{23} \\
\eta_{03} & \eta_{13} & \eta_{23} & \eta_{33} 
\end{pmatrix}.
\end{equation}

Since both Higgs doublets carry the same quantum numbers
we may also consider the unitarily mixed fields
\begin{equation}
\label{eq11}
\begin{pmatrix} \varphi_1(x) \\
                \varphi_2(x) \end{pmatrix}
\to
\begin{pmatrix} \varphi'_1(x) \\
                \varphi'_2(x) \end{pmatrix}
= U
  \begin{pmatrix} \varphi_1(x) \\
                  \varphi_2(x) \end{pmatrix}\;,
\end{equation}
with $U= (U_{ij}) \in \text{U(2)}$. For the bilinears
a basis, or Higgs-family, transformation~\eqref{eq11} of
the fields corresponds
to a SO(3) rotation given by
\begin{equation}
\label{eqbasistfK}
\begin{split}
K_0(x) &\to K'_0(x) = K_0(x),\\
\tvec{K}(x) &\to \tvec{K}'(x) = R(U)\; \tvec{K}(x)\;.
\end{split}
\end{equation}
Here $R(U)$ is obtained from
\begin{equation}
\label{eq13}
U^\dagger \sigma^a U = R_{ab}(U)\,\sigma^b.
\end{equation}
We note that every proper rotation matrix~$R \in \text{SO(3)}$ is a rotation about
an axis and can be represented, in a suitable basis, as
\begin{equation}
\label{eqrot}
R_\alpha = 
\begin{pmatrix}
\cos (\alpha)  & - \sin (\alpha) & 0 \\
\sin (\alpha) & \cos (\alpha) & 0 \\
0 & 0 & 1
\end{pmatrix},
\end{equation}
where~$\alpha$ is the angle of rotation.

We shall also consider
generalized CP~(GCP)
transformations~\cite{Lee:1966ik,Ecker:1981wv,Ecker:1983hz,Bernabeu:1986fc,Ecker:1987qp,Neufeld:1987wa,Lavoura:1994fv, Botella:1994cs}, where
\begin{equation}
\label{eqX}
\varphi_i(x) \rightarrow U_{ij} \varphi_j^*(x') , 
\qquad i,j=1,2\;,
\quad x=(x^0, \tvec{x}) ,
\quad x'=(x^0, -\tvec{x})
\end{equation}
with $U=(U_{ij}) \in \text{U(2)}$.
Note that the ordinary CP transformation is
the special case of $U=\unitmatrix_2$ in~\eqref{eqX}.
In $K$ space the generalized CP transformations~\eqref{eqX} correspond 
to the improper rotations~\cite{Maniatis:2007vn,Ferreira:2009wh}
\begin{equation}
\label{eqGCPK}
\begin{split}
K_0(x) &\to K_0(x') ,\\
\tvec{K}(x) &\to \tvec{K}'(x') = \bar{R}(U) \; \tvec{K}(x') .
\end{split}
\end{equation}
Here
\begin{equation}
\bar{R}(U) = R(U) \bar{R}_2
\end{equation}
with~$\bar{R}_2$ the matrix for reflection
on the 1--3 plane. We define the matrices
$\bar{R}_j$ ($j=1,2,3$) for the reflections
on the coordinate planes in~$K$ space as
\begin{equation}
\label{eqreflect}
\bar{R}_1 = \diag (-1,1,1), \qquad 
\bar{R}_2 = \diag (1,-1,1), \qquad 
\bar{R}_3 = \diag (1,1,-1) .
\end{equation}
Here and in the following proper rotation matrices
will be denoted by $R$, $R_\alpha$, etc., improper
rotation matrices by $\bar{R}$, $\bar{R}_j$,
$\bar{R}_\alpha$, etc.
By a suitable basis choice
we can always arrange that the improper rotation matrix~$\bar{R}(U)$
has the form
\begin{equation}
\label{GCPK}
\bar{R}_\alpha=
\begin{pmatrix}
\cos (\alpha) & -\sin (\alpha) & 0\\
\sin (\alpha) &  \cos (\alpha) & 0\\
0 & 0 & -1\\
\end{pmatrix},
\qquad \text{with } 0 \le \alpha \le \pi \;.
\end{equation}
Note that for $\alpha=0$ we get the GCP transformation
corresponding to a reflection on the 1--2 plane in~$K$ space
($\bar{R}(U)=\bar{R}_3$)
accompanied by the space-time transformation $x \to x'$. 
A basis transformation~\eqref{eqbasistfK}
exchanging the 2 and 3 axes in~$K$ space shows
that this is equivalent to the standard CP
transformation where~$\bar{R}(U)=\bar{R}_2$
in~\eqref{eqGCPK}.
For more details on GCPs in THDMs see~\cite{Maniatis:2007vn,Ferreira:2009wh,
Ferreira:2010hy}.

Finally we recall from~\cite{Maniatis:2006fs} that 
a transformation~\eqref{eqbasistfK}
in $K$ space with $R \in \text{SO(3)}$ always
corresponds to a field transformation~\eqref{eq11}
which is unique up to gauge transformations.
Similarly, a $K$-space transformation
\eqref{eqGCPK} with $\bar{R} \in \text{O(3)}$,
$\det (\bar{R})=-1$,
always corresponds to a GCP transformation~\eqref{eqX}
of the fields which is unique up to
gauge transformations.

%
\section{Symmetry classes and symmetry groups}

The general THDM potential has~14 parameters; 
see~\eqref{connect}.
Considering only the scalar sector we can make
a basis change as in~\eqref{eq11}, \eqref{eqbasistfK}
to diagonalise~$E=\diag (\mu_1, \mu_2, \mu_3)$,
thereby reducing the number of parameters to~11.
One may want to further reduce this number by imposing
symmetries. This can be Higgs-family or GCP
symmetries. A Higgs-family
transformation~\eqref{eq11}, \eqref{eqbasistfK}
is a symmetry of the potential if and only if 
the parameters~\eqref{connect}
satisfy
\begin{equation}
\label{eqR}
R(U) \tvec{\xi} = \tvec{\xi} , \quad
R(U) \tvec{\eta} = \tvec{\eta} , \quad
R(U) E R^\trans(U) = E .
\end{equation}
A GCP transformation~\eqref{eqX}, \eqref{eqGCPK} is 
a symmetry if and only if
\begin{equation}
\label{eqRbar}
\bar{R}(U) \tvec{\xi} = \tvec{\xi} , \quad
\bar{R}(U) \tvec{\eta} = \tvec{\eta} , \quad
\bar{R}(U) E \bar{R}^\trans(U) = E .
\end{equation}

In~\cite{Ivanov:2007de} the possible symmetry classes
of THDMs were derived, however, only 
potentials which are stable in the 
{\em strong sense} were considered.
Here we define, as in \cite{Maniatis:2006fs}, a potential
to be stable in the strong sense if stability is guaranteed
by the quartic field terms alone and in the {\em weak sense}
if it is guaranteed only after inclusion of the quadratic
field terms in~\eqref{eqV} respectively \eqref{eqVK}.
A potential being bounded from below but having
directions in field space where it does not grow
indefinitely for the fields going to infinity has only
marginal stability.
In all other cases the potential is unstable.
In~\cite{Ferreira:2010hy} the symmetry classes of the THDMs were further
studied and also softly broken symmetries were considered.

We give in Table~\ref{tabsymmgroup} the maximal
symmetry group for each symmetry class and the corresponding
constraints on the potential~\eqref{eqVK}.
Note that in Table~\ref{tabsymmgroup} the classes are defined to be mutually
exclusive, that is, we assign a THDM to a
certain class if it has the corresponding
group~$\mathbbm{G}$ (up to trivial equivalences)
as symmetry group and {\em not} a bigger one.
If the parameters of a THDM potential are not
satisfying any of the constraints of Table~\ref{tabsymmgroup},
the theory has no symmetry group except the trivial one, 
that is, the unit transformation. 
In appendix~\ref{appA} we present a derivation of
these symmetry classes and groups where, as mentioned
above, we do not use any assumptions on the
stability of the potential~\eqref{eqV}, \eqref{eqVK}.
The methods explained in apppendix~\ref{appA} also give
an easy practical recipe for finding out if a
THDM potential has a symmetry and which one this is.
The symmetry relations as given in Table~\ref{tabsymmgroup}
will be used in section~\ref{secren} for the discussion of
the renormalisation in specific THDMs.
\begin{table*}[t!]
\centering
\begin{tabular}{lllc}
\hline
\multicolumn{2}{l}{symmetry class and group~$\mathbbm{G}$} &
constraints on $\tvec{\xi}$ and $\tvec{\eta}$ &
constraints on $E$
\\ 
\hline
$\mathbb{Z}_2$ &
$\{ \unitmatrix_3, \bar{R}_1, \bar{R}_2, \bar{R}_1 \bar{R}_2 \}$ &
$\left\{
\begin{array}{l}
\tvec{\xi} \times \tvec{e}_3 =0,\; 
\tvec{\eta} \times \tvec{e}_3 =0,\;
(\tvec{\xi},\tvec{\eta})\neq (0,0)
\\ [1pt]
\xi_1=0,\; \eta_1=0,\; \tvec{\xi}\times\tvec{\eta}=0,\;
(\tvec{\xi},\tvec{\eta})\neq (0,0)
\end{array} \right.$
&
$\begin{array}{r}
\mbox{all  $\mu_i$  different}\\
\mbox{$\mu_1 \neq \mu_2 =\mu_3$}
\end{array}$
\\ 
\hline
U(1) &
$\{ R_{2 \theta}, R_{2 \theta} \bar{R}_2 \}$ 
&
$\left\{
\begin{array}{l}
\tvec{\xi} \times \tvec{e}_3 =0,\; 
\tvec{\eta} \times \tvec{e}_3 =0,\;
(\tvec{\xi},\tvec{\eta})\neq (0,0)
\\
\tvec{\xi}\times\tvec{\eta}=0,\;
(\tvec{\xi},\tvec{\eta})\neq (0,0)
\end{array}\right.$
&
$\begin{array}{r}
\mu_1 = \mu_2 \neq \mu_3\\
\mu_1 = \mu_2 =\mu_3
\end{array}$
\\ 
\hline
SO(3) &
$\{ R, R \bar{R}_2 \}$ 
&
\quad $\tvec{\xi} = 0,\; 
\tvec{\eta} = 0$
& \;$\mu_1 = \mu_2 =\mu_3$
\\  
\hline
CP1 &
$\{ \unitmatrix_3, \bar{R}_2 \}$
&
$\left\{ 
\begin{array}{l}
\xi_2 = 0$, $\eta_2=0,\;
(\xi_1,\eta_1) \neq (0,0),\; 
(\xi_3,\eta_3) \neq (0,0)
\\
\xi_2 = 0$, $\eta_2=0,\;
\tvec{\xi}\times\tvec{\eta} \neq 0 
\\
(\xi_3,\eta_3) \neq (0,0),\;
(\tvec{\xi}\times\tvec{\eta}) \cdot \tvec{e}_3 = 0,\;
(\tvec{\xi}-\xi_3 \tvec{e}_3, \tvec{\eta}-\eta_3 \tvec{e}_3) \neq (0,0)
\\
\tvec{\xi}\times\tvec{\eta} \neq 0
\end{array} \right.$
&
\begin{small}
$\begin{array}{r}
\text{all } \mu_j \text{ different} 
\\
\mu_1 = \mu_3 \neq \mu_2 
\\
\mu_1 = \mu_2 \neq \mu_3 
\\
\mu_1 = \mu_2 = \mu_3 
\end{array}$
\end{small}
\\
\hline 
\multirow{2}{*}{CP2} &
$\{ \unitmatrix_3, \bar{R}_1, \bar{R}_2,\bar{R}_3,
\bar{R}_1 \bar{R}_2, \bar{R}_2 \bar{R}_3,$
&
\multirow{2}{*}{\quad $\tvec{\xi}=0$, $\tvec{\eta}=0$} 
&
\multirow{2}{*}{\; all $\mu_j$ different}
\\
&
$\bar{R}_1 \bar{R}_3,
\bar{R}_1 \bar{R}_2 \bar{R}_3 =-\unitmatrix_3 \}$ & &\\
\hline
CP3 &
$\{R_{2	\theta}, R_{2 \theta} \bar{R}_2, R_{2 \theta} \bar{R}_3 \}$ 
&
\quad $\tvec{\xi}=0$, $\tvec{\eta}=0$ 
&
\; $\mu_1 = \mu_2 \neq \mu_3$ \\
\hline
\end{tabular}
\caption{\label{tabsymmgroup} \small The symmetry classes,
groups~$\mathbbm{G}$, and the corresponding
constraints on the scalar-potential parameters.
The eigenvalues of~$E$ are denoted by~$\mu_j$ with $j=1,2,3$ and
the vector $\tvec{e}_3 = (0,0,1)^\trans$.
$\mathbbm{G}$ is the symmetry group defining the class.
The matrices $R_{2 \theta} \in \text{SO(3)}$ with $0 \le\theta < \pi$
are defined in~\eqref{eqA11} and \eqref{eqrot} with
$\alpha=2\theta$,
the reflection matrices~$\bar{R}_j$ in~\eqref{eqreflect}.}
\end{table*}

We emphasize that in Table~\ref{tabsymmgroup} we give the {\em exact}
conditions for the parameters of the scalar potential 
to have the symmetry group~$\mathbbm{G}$ as listed
and {\em not} a bigger one.
The elements of~$\mathbbm{G}$ give the corresponding
transformations in $K$ space.
For proper rotations these are Higgs basis
transformations; see eq.~\eqref{eqbasistfK}, for
improper rotations these are generalized CP transformations;
see~eq.~\eqref{eqGCPK}.
Of course, a group $\mathbbm{G}$ of a symmetry class
may contain the groups of other classes as subgroups,
as is obvious from Table~\ref{tabsymmgroup}.
For instance, the group O(3) contains all
other groups as subgroups
and, clearly, the potential of the SO(3)
symmetry class has all other symmetries as well.
The numbering of the eigenvalues of~$E$ 
in Table~\ref{tabsymmgroup} is - without loss
of generality - chosen conveniently,
in order to give the same invariance group
$\mathbbm{G}$ 
and not an equivalent one
for all subclasses of one class.
For the cases of degenerate eigenvalues of~$E$
it is understood that a convenient choice 
of basis in the degenerate subspaces
gives the groups $\mathbbm{G}$ as listed.
Other choices of bases give equivalent groups.

In Table~\ref{tabsymmgroup} we have listed subclasses
for $\mathbb{Z}_2$, U(1), and CP1.
These are distinguished by the
degeneracies of the eigenvalues~$\mu_j$ and for 
the CP1 case {\em also} by relations for
$\tvec{\xi}$ and $\tvec{\eta}$. These
subclasses of a class correspond to the same
symmetry group $\mathbbm{G}$ and
therefore lead to no new symmetry classes.
Under renormalisation only the groups
$\mathbbm{G}$ will be preserved. That is,
the subclasses of one class will not
be invariant under renormalisation but 
will mix among each other. 
Considering the theory of the two Higgs-boson
doublets alone the renormalisation of the potential
parameters can not lead from one symmetry class
to another one. 
If we start, for instance, with a theory of
the CP2 class where all $\mu_i$ are different
we can not come by renormalisation to
the CP3 or SO(3) classes where two, respectively
all three, of the $\mu_i$'s are equal. We shall
elaborate on this point below in section~\ref{secren}
in connection with the renormalisation in the
MCPM which is a complete theory including
fermions and bosons.

The elements of the various symmetry groups 
in Table~\ref{tabsymmgroup} are listed
according to their action in~$K$ space;
see~\eqref{eqbasistfK}, \eqref{eqGCPK}.
For completeness we list in appendix~\ref{appA}
also the corresponding transformations for 
the fields.

%
\section{Renormalisation of the dimension four couplings in the MCPM}
\label{secren}

In this section we consider the 
renormalisation-group equations (RGEs)
for the dimension four couplings in
the maximally CP symmetric model~(MCPM)
as constructed and studied 
in~\cite{Maniatis:2007de,Maniatis:2009vp,Maniatis:2009by,Maniatis:2010sb}.
In the MCPM the Higgs potential parameters~\eqref{connect},
in a diagonal basis of
the matrix~$E$, have to fulfill 
\begin{equation}
\label{MCPMpot}
\tvec{\xi}=0, \qquad \tvec{\eta}=0,\qquad E=\diag (\mu_1, \mu_2, \mu_3).
\end{equation}
In conventional notation of
the Higgs potential~\eqref{eqV} this corresponds
to the constraints 
\begin{equation}
\label{MCPMpotphi}
m_{12}^2=0,\qquad m_{11}^2=m_{22}^2,\qquad 
\lambda_1=\lambda_2,\qquad \im (\lambda_5)=0,\qquad
\lambda_6= \lambda_7=0. 
\end{equation}
Without loss of generality we can assume
\begin{equation}
\label{eqmuord}
\mu_1 \ge \mu_2 \ge \mu_3 .
\end{equation}

From Table~\ref{tabsymmgroup} we see that the Higgs potential
satisfying~\eqref{MCPMpot} can be in the symmetry
classes CP2, CP3, or SO(3). As shown in~\cite{Maniatis:2007de},
stability, the correcte electroweak symmetry breaking~(EWSB),
and absence of zero-mass charged Higgs bosons require and
are guaranteed by
\begin{gather}
\begin{split}
\label{eqCPicond}
\eta_{00} >0 \;,\qquad
\mu_i + \eta_{00} > 0 \quad {\text{for  }} i=1,2,3 \;,\qquad
\xi_0 <0\;,\qquad
\mu_3 <0 .
\end{split}
\end{gather}
In the MCPM there are five physical Higgs bosons, three neutral ones,
$\rho'$, $h'$, $h''$, and a charged pair, $H^\pm$. Their
squared masses in terms of the model parameters are,
at tree level,
\begin{equation}
\label{eq21}
m_{\rho'}^2\;= 2 v_0^2 (\eta_{00}+\mu_3)\;,\quad
m_{h'}^2\;= 2 v_0^2 (\mu_1 - \mu_3)\;,\quad
m_{h''}^2\;= 2 v_0^2 (\mu_2 - \mu_3)\;,\quad
m_{H^\pm}^2= 2 v_0^2 (- \mu_3)\;.
\end{equation}
Here
\begin{equation}
v_0 = \sqrt{ \frac{ -\xi_0 }{ \eta_{00} + \mu_3 }} \approx 246~\text{GeV}
\end{equation}
is the standard vacuum-expectation value.
Requiring now also absence of zero-mass neutral
Higgs bosons and absence of mass degeneracy between
$h'$ and $h''$ leads to
\begin{equation}
\label{eqmuordstrong}
\mu_1 > \mu_2 > \mu_3,
\end{equation}
replacing the weaker condition~\eqref{eqmuord}.
From Table~\ref{tabsymmgroup} we see that
we are dealing now with potentials in the CP2
symmetry class with the corresponding
symmetry group~$\mathbbm{G}$ as listed
there. The main point of the MCPM is that
the symmetry group of the CP2 class is required
to be respected also by the complete Lagrangian,
including the fermions, the gauge-boson, and the
Yukawa sectors.
It was shown in~\cite{Maniatis:2007de} that
with this requirement a coupling of the two
Higgs-boson doublets
to only one fermion family
is not possible with non-vanishing Yukawa couplings. 
However, with a coupling of the two Higgs-boson doublets
to two fermion families
this is indeed possible with one fermion family acquiring masses
and the other remaining massless. 
With a third fermion family 
kept uncoupled to the Higgs-boson doublets this model 
gives very roughly what we observe in Nature:
two rather light fermion families and one
very heavy (the third) fermion family.

The complete Lagrangian of the MCPM is recalled in App.~\ref{appLag}.
The parameters of the MCPM are as follows.
\begin{itemize}
\item Higgs potential parameters
\begin{equation}
\xi_0,\quad \eta_{00},\quad \mu_1,\quad \mu_2,\quad \mu_3 .
\end{equation}
\item Yukawa sector coupling constants
\begin{equation}
c_\tau,\quad c_t,\quad c_b
\end{equation}
related to the third-fermion-family masses
\begin{equation}
\label{tbmasses}
m_\tau = c_\tau \frac{v_0}{\sqrt{2}}\;,\qquad
m_t = c_t \frac{v_0}{\sqrt{2}}\;,\qquad
m_b = c_b \frac{v_0}{\sqrt{2}} .
\end{equation}
\item Gauge couplings
\begin{equation}
g_1,\quad g_2,\quad g_3
\end{equation}
of the gauge groups $\text{U(1)}_Y$, $\text{SU(2)}_L$, and $\text{SU(3)}_C$,
respectively.
\end{itemize}

Let us now proceed and consider the
one-loop RGEs
in this model.
The one-loop RGEs 
for the couplings of 
the dimension-four terms in
any renormalisable gauge theory
are given in~\cite{Cheng:1973nv}.
The RGEs given there apply to the
{\em deep Euclidean region} where coupling
terms of dimension two can be neglected.
Also shifts of scalar fields to give them
zero vacuum expectation value after EWSB
are irrelevant there. 
For the quartic Higgs-potential couplings
$\lambda_{1,2,3,4,5,6,7}$ including the $\text{U(1)}_Y$ and $\text{SU(2)}_L$ 
gauge interactions with couplings $g_1$ and $g_2$, respectively,
taking also the Yukawa couplings~\eqref{eqYuk} into account
we find for the MCPM from the results of~\cite{Cheng:1973nv}
\begin{equation}
\begin{split}
\label{rgelambda}
8 \pi^2 \frac{d \lambda_1}{dt}  =&
6 \lambda_1^2 + 2 \lambda_3^2 
+ 2 \lambda_3 \lambda_4 + \lambda_4^2 + \lambda_5^2
- \lambda_1 \left( \frac{3}{2} g_1^2 + \frac{9}{2} g_2^2 \right) 
+ \frac{3}{8} g_1^4 + \frac{3}{4} g_1^2 g_2^2 + \frac{9}{8} g_2^4
+ 2 \lambda_1 (c_\tau^2+c_b^2+c_t^2) - 2 (c_\tau^4+c_b^4+c_t^4),
\\
8 \pi^2 \frac{d \lambda_3}{dt} =& 2\lambda_1(3 \lambda_3 
+ \lambda_4) + 2 \lambda_3^2 + \lambda_4^2 + \lambda_5^2  
- \lambda_3 \left( \frac{3}{2} g_1^2 + \frac{9}{2} g_2^2 \right) 
+ \frac{3}{8} g_1^4 - \frac{3}{4} g_1^2 g_2^2 + \frac{9}{8} g_2^4
+ \lambda_3 (c_\tau^2 + c_b^2 +c_t^2),
\\ 
8 \pi^2 \frac{d \lambda_4}{dt} =& 2 \lambda_1 \lambda_4 
+ 4 \lambda_3 \lambda_4 + 2 \lambda_4^2 + 4 \lambda_5^2  
- \lambda_4 \left( \frac{3}{2} g_1^2 + \frac{9}{2} g_2^2 \right) 
+ \frac{3}{2} g_1^2 g_2^2
+ \lambda_4 (c_\tau^2+c_b^2 +c_t^2),
\\
8 \pi^2 \frac{d \lambda_5}{dt} =& \lambda_5  \left (2 \lambda_1 
+ 4 \lambda_3 + 6 \lambda_4 \right)
- \lambda_5 \left( \frac{3}{2} g_1^2 + \frac{9}{2} g_2^2 \right)
+ \lambda_5 (c_\tau^2+c_b^2 +c_t^2),
\\
\frac{d \lambda_2}{dt} =& \frac{d \lambda_1}{dt},
\qquad
\frac{d \lambda_6}{dt}= \frac{d \lambda_7}{dt}=0 .
\end{split}
\end{equation}
Here $t=\ln (M/M_0) $ with $M$ the 
mass scale of the renormalisation point and~$M_0$
a convenient reference scale, for instance, $M_0= 1$~TeV.
The RGEs of the $\lambda$'s can easily be translated to $K$ space.
For the generic THDM Higgs potential this was done in~\cite{Ma:2009ax}.
In the case of the MCPM we have to extend these
RGEs by including the Yukawa interactions~\eqref{eqYuk}.
From the RGEs for the parameters of the generic Higgs potential
as given in~\cite{Ma:2009ax} we can check that the diagonality
of the matrix~$E$ and $\tvec{\eta}=0$, see~\eqref{MCPMpot},
are preserved under one-loop renormalisation in the MCPM. This 
must be so, since this is guaranteed by the symmetry 
group~$\mathbbm{G}$ of the CP2 class; see Table~\ref{tabsymmgroup}.
Here we find
from~\eqref{connect}, \eqref{Kquart}, and \eqref{rgelambda},
\begin{equation}
\begin{split}
\label{rgeK}
8 \pi^2 \frac{d \eta_{00}}{dt} =& 4 \eta_{00}^2 + \eta_{00} (\eta_{11} 
+ \eta_{22} + \eta_{33}) + \eta_{11}^2 + \eta_{22}^2 + \eta_{33}^2 
- \eta_{00} \left( \frac{3}{2} 
g_1^2 + \frac{9}{2} g_2^2 \right) + \frac{3}{4} g_1^4 + \frac{9}{4} g_2^4\\
& + (\frac{3}{2} \eta_{00} + \frac{1}{2} \eta_{33}) (c_\tau^2+c_b^2+c_t^2)
- \frac{1}{2} (c_\tau^4+c_b^4+c_t^4),
\\ 
8 \pi^2 \frac{d \eta_{11}}{dt} =& \eta_{11} \left( 3 \eta_{00} + 3 \eta_{11} - 
\eta_{22} - \eta_{33} - \frac{3}{2} g_1^2 - \frac{9}{2} g_2^2 
+c_\tau^2 + c_b^2 + c_t^2 \right) + 
\frac{3}{2} g_1^2 g_2^2, 
\\
8 \pi^2 \frac{d \eta_{22}}{dt} =& \eta_{22} \left( 3 \eta_{00} - \eta_{11} + 
3 \eta_{22} - \eta_{33} - \frac{3}{2} g_1^2 - \frac{9}{2} g_2^2 
+ c_\tau^2 + c_b^2 + c_t^2 \right) + 
\frac{3}{2} g_1^2 g_2^2,
\\
8 \pi^2 \frac{d \eta_{33}}{dt} =& \eta_{33} \left( 3 \eta_{00} - \eta_{11} - 
\eta_{22} + 3 \eta_{33} - \frac{3}{2} g_1^2 - \frac{9}{2} g_2^2 \right)
+ \frac{3}{2} g_1^2 g_2^2\\
& + (\frac{1}{2} \eta_{00} + \frac{3}{2} \eta_{33}) (c_\tau^2+c_b^2+c_t^2)
- \frac{1}{2} (c_\tau^4+c_b^4+c_t^4).
\end{split}
\end{equation}
As mentioned above these RGEs apply in the
deep Euclidean region.

Let us now discuss the evolution of the differences of
the eigenvalues of~$E$:
\begin{equation}
\mu_1-\mu_2 , \qquad \mu_2-\mu_3 .
\end{equation}
From~\eqref{rgeK} we find
\begin{equation}
\label{eqrgediff}
\begin{split}
8 \pi^2 \frac{d}{dt} (\mu_1 - \mu_2) =& 
(\mu_1 - \mu_2)
\big[ 3 (\eta_{00}+\mu_1+\mu_2 )
- \mu_3 -\frac{3}{2} g_1^2 - \frac{9}{2} g_2^2
+c_\tau^2+c_b^2 + c_t^2 \big],\\
8 \pi^2 \frac{d}{dt} (\mu_2 - \mu_3) =& 
(\mu_2 - \mu_3)
\big[ 3 (\eta_{00}+\mu_2+\mu_3 )
- \mu_1 -\frac{3}{2} g_1^2 - \frac{9}{2} g_2^2
+c_\tau^2+c_b^2 + c_t^2 \big]\\
& -\frac{1}{2} (\eta_{00}+\mu_3)(c_\tau^2+c_b^2+c_t^2)
+\frac{1}{2} (c_\tau^4+c_b^4+c_t^4) .
\end{split}
\end{equation}

Suppose now that we start at $M_0=1~$TeV, corresponding
to $t=0$, with the conditions~\eqref{eqmuordstrong}.
We have then, in particular,
\begin{equation}
\big[ \mu_1(t) - \mu_2(t) \big]\big|_{t=0} > 0 .
\end{equation}
From~\eqref{eqrgediff} we can see that the
one loop RGEs preserve this property as long as all couplings stay
finite. Indeed, suppose that for $0 \le t \le t_1$ we have
\begin{equation}
\big| 3 (\eta_{00}+\mu_1 + \mu_2) - \mu_3
-\frac{3}{2} g_1^2 - \frac{9}{2} g_2^2 + c_\tau^2 + c_b^2 + c_t^2 \big|
\le 8 \pi^2 C ,
\end{equation}
where $C>0$ is a constant. We get then from~\eqref{eqrgediff}
\begin{equation}
8 \pi^2 \frac{d}{dt} \ln (\mu_1 - \mu_2)
= 
3 (\eta_{00}+\mu_1+\mu_2 )
- \mu_3 -\frac{3}{2} g_1^2 - \frac{9}{2} g_2^2
+c_\tau^2 + c_b^2 + c_t^2 ,
\end{equation}
\begin{equation}
- C \le \frac{d}{dt} \ln (\mu_1 - \mu_2) \le C ,
\end{equation}
\begin{equation}
\label{eqmuevol}
e^{-C t} \le 
\frac{\mu_1(t) - \mu_2(t)}{\mu_1(0) - \mu_2(0)} \le e^{ C t},
\qquad \text{for } 0 \le t \le t_1 .
\end{equation}
Thus, $\mu_1(t) - \mu_2(t)$ stays positive for $0 \le t \le t_1$.
A similar argument applies for the evolution to negative~$t$ values.
Hence, $\mu_1(t) - \mu_2(t)$ can not change sign as long as the
theory parameters stay finite.

The analogous result for~$\mu_2 - \mu_3$ can {\em not}
be derived in the same way from~\eqref{eqrgediff}.
This is due to the terms not proporotional to $\mu_2 - \mu_3$
on the r.h.s of \eqref{eqrgediff}.
But in the pure scalar theory, that is, if we set
$g_1=g_2=0$ and $c_b=c_t=c_\tau=0$ we can again
derive the analogue of~\eqref{eqmuevol}.

We conclude that the one loop RGEs preserve 
$\mu_1 > \mu_2$ but, in the full theory,
not necessarily $\mu_2 > \mu_3$
If now for some $t$-value $t_0$ we have
$\mu_2(t_0)=\mu_3(t_0)$ we have for the Higgs potential
a higher symmetry, here the CP3 symmetry, where
two eigenvalues of~$E$ are equal; see Table~\ref{tabsymmgroup}.
But for the {\em full theory} this CP3 symmetry is {\em not}
realised. Thus, in the full theory the RGEs can lead
to renormalisation scales~$M$ where the Higgs potential alone
shows a higher symmetry than the full theory. Of course,
only the symmetry of the full theory is relevant for physics.

%
\section{Oblique parameters in the MCPM}
\label{secob}
The oblique parameters~$S$, $T$, and $U$ denote 
certain combinations of
self-energies of 
the electroweak gauge bosons with respect
to any new contributions compared to the SM~\cite{Peskin:1990zt}.
In any model beyond the SM the oblique
parameters can be computed and compared to the
electroweak precision data~\cite{Nakamura:2010zzi}
which require:
\begin{equation}
\label{STUex}
S=0.01 \pm 0.10, \qquad
T=0.03 \pm 0.11, \qquad
U=0.06 \pm 0.10.
\end{equation}
For the case of the general THDM the oblique parameters
have been computed in~\cite{Froggatt:1991qw,Haber:2010bw}.

\begin{figure}
\includegraphics[height=0.9\textheight]{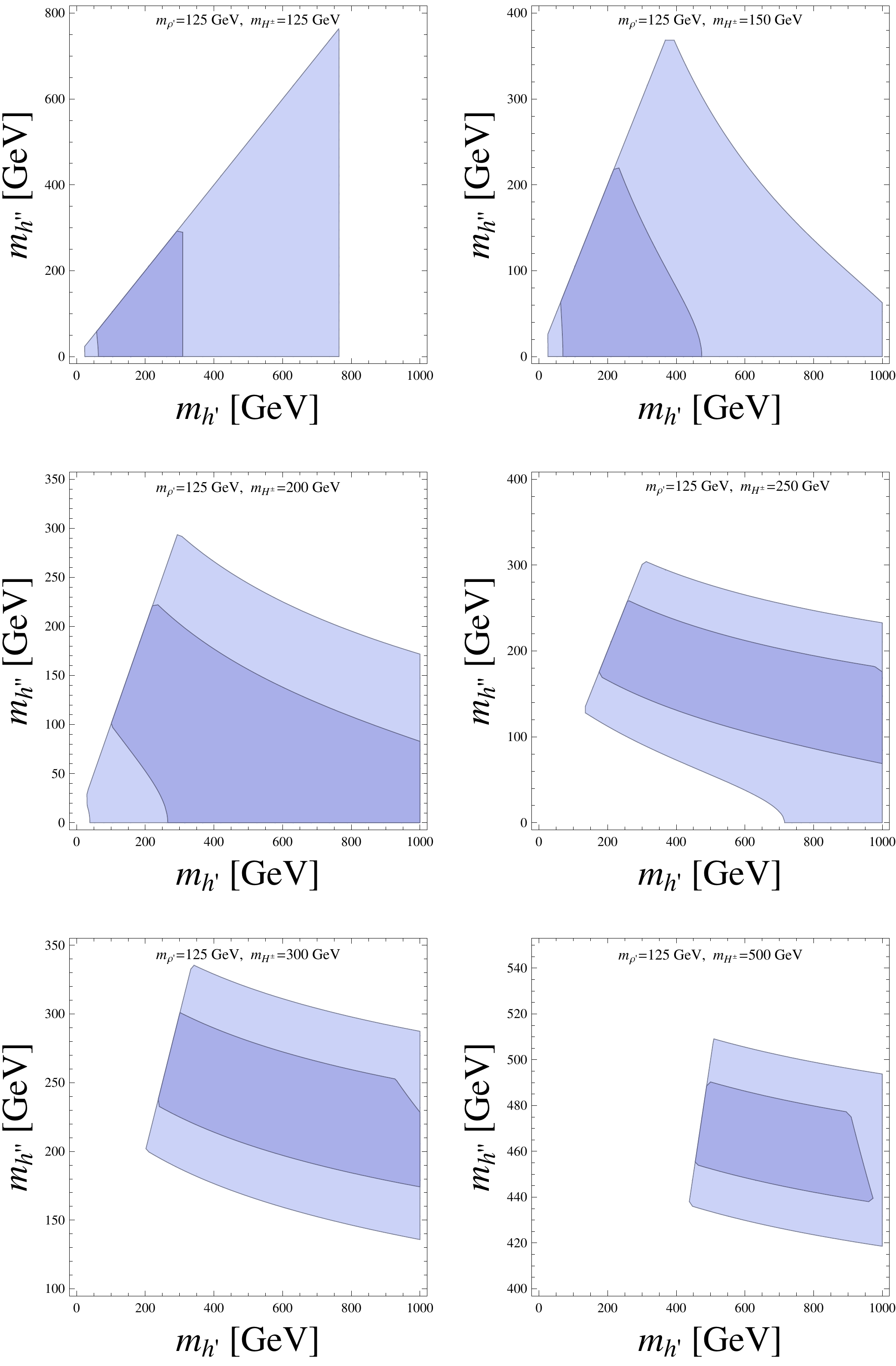}
\caption{\label{figob}
The allowed regions for the Higgs-boson masses $m_{h'}$
and $m_{h''}$ corresponding to the
1-$\sigma$ (dark) and 2-$\sigma$ (bright) uncertainties 
in the measured oblique parameters $S$, $T$, and $U$~\eqref{STUex}.
The contours are shown for a fixed value of the SM-like Higgs-boson mass,
$m_{\rho'}=125$~GeV, and different choices of the charged-Higgs-boson
mass $m_{H^\pm}$ as indicated within the plots.}
\end{figure}
\begin{figure}
\includegraphics[height=0.9\textheight]{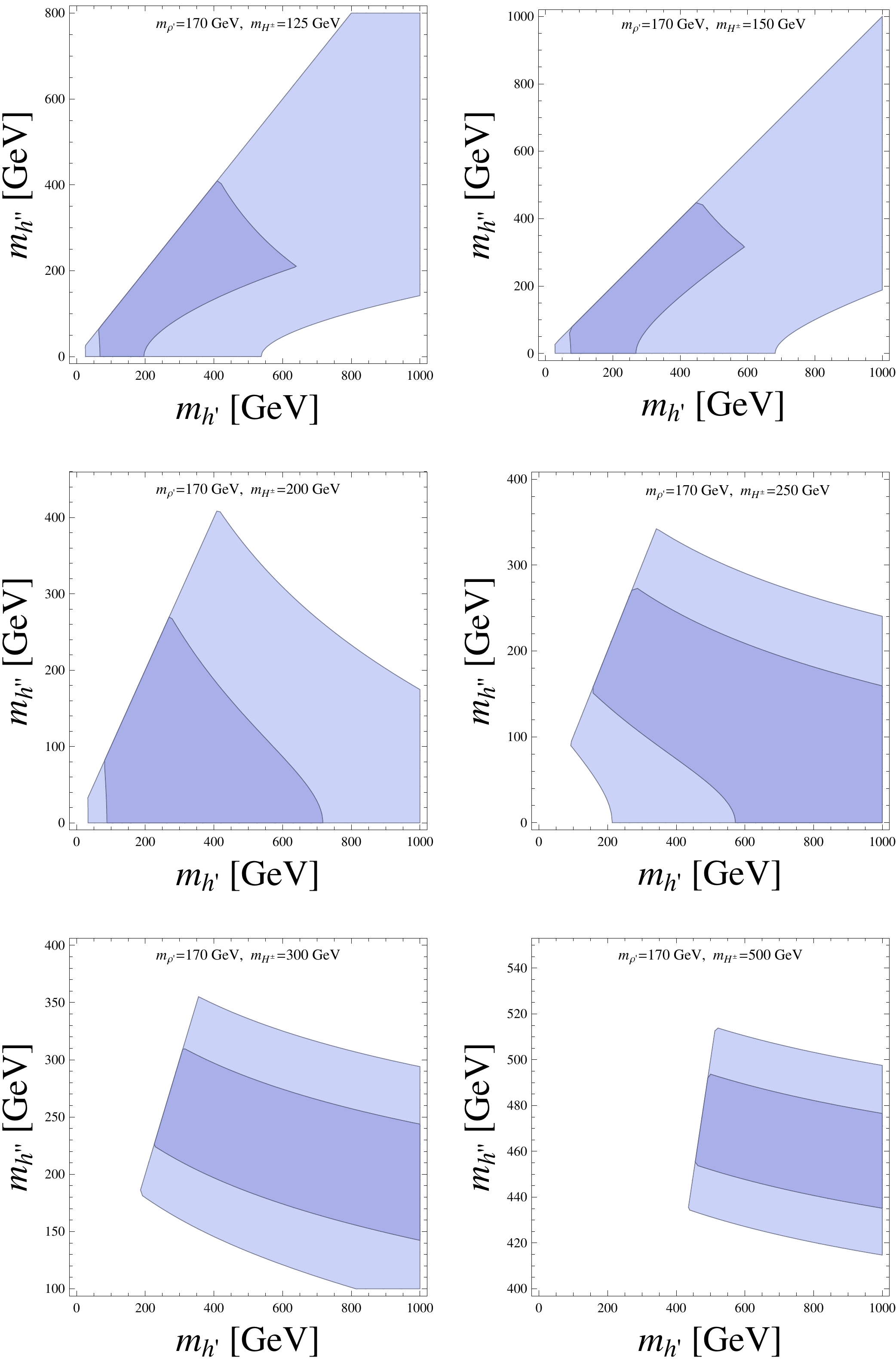}
\caption{\label{figob2}
Same as
in Fig.~\ref{figob} but with
the SM-like Higgs-boson mass fixed to $m_{\rho'}=170$~GeV.}
\end{figure}

We shall now derive the predictions
for the oblique parameters in the MCPM.
In the MCPM the Yukawa couplings are
completely fixed and the only free parameters we encounter
in the calculation of the oblique parameters are
the Higgs-boson masses
$m_{\rho'}$, $m_{h'}$, $m_{h''}$, and $m_{H^\pm}$.
Here $\rho'$ and $h'$ are the CP-even and
$h''$ is the CP-odd Higgs boson and
$H^\pm$ denotes the pair of charged Higgs bosons.
In Figure~\ref{figob} we show the contour plots
for the 1-$\sigma$ (dark) and 2-$\sigma$ (bright) deviations of the oblique
parameters from the electroweak precision
data~\eqref{STUex} in the~$m_{h'}$--$m_{h''}$ plane. 
The mass of the
SM-like Higgs-boson~$\rho'$ is fixed
to $m_{\rho'} =$~125~GeV.
The charged-Higgs-boson
mass $m_{H^\pm}$ is set to different values
in the range of
125-500~GeV in the various plots. 
In Figure~\ref{figob2} we show analogous
plots but for a mass of the SM-like Higgs
boson~$\rho'$ of $m_{\rho'} =$~170~GeV.
Note that we have always 
$m_{h'} > m_{h''}$ in the MCPM which is the reason that
there are no allowed regions of
parameter space above the diagonal of equal 
masses $m_{h'}=m_{h''}$ in Figures~\ref{figob}
and~\ref{figob2}

We see from Figures~\ref{figob}
and \ref{figob2} that there are large regions
for the masses of the Higgs bosons $h'$, $h''$, and
$H^\pm$ where the electroweak constraints~\eqref{STUex}
are satisfied. The allowed regions for these masses,
up to several hundred GeV,
are very reasonable.
The CP odd extra Higgs boson~$h''$ could even be
below~100~GeV in mass. But then it would be necessary
to study all other experimental constraints for such
a low-mass boson.
Furthermore, we see from Figures~\ref{figob}
and \ref{figob2} that with increasing masses
of~$\rho'$ and $H^\pm$ also the allowed domains for the
masses of the Higgs bosons $h'$ and $h''$ shift to
higher mass values.

%
\section{Conclusions}
\label{seccon}

In this paper we started with briefly reviewing the
bilinear formalism which turns out to be quite
powerful for the study of the THDM. We
have discussed the classification of
the possible symmetry classes 
without any assumption on the stability
type of the THDM potential.
We have given a practical and 
easily usable method how to determine
the symmetry class of a given THDM
Higgs potential. We have
defined the
symmetry classes to be mutually exclusive;
see Table~\ref{tabsymmgroup}.
We have also given
the symmetry
group~$\mathbbm{G}$ corresponding to each
symmetry class. We have focussed on one
of these symmetry classes, denoted by CP2, in some detail.
The CP2 symmetric THDM has a number of
simultaneous CP invariances.
As shown in \cite{Maniatis:2007de}
the extension of the
CP symmetries of the potential
to the
Yukawa interactions leads in a straightforward way
to the so-called maximally CP-symmetric model (MCPM).
In this model
the Yukawa couplings are completely fixed.
We have
studied the renormalisation-group equations of
the quartic Higgs-potential parameters in this model.
We have found that the symmetries of this model
are preserved by the RGEs, as it has to be.

The MCPM has a hierarchy of
quartic couplings $\mu_1>\mu_2>\mu_3$.
We have shown that considering the theory of
the Higgs bosons alone
this hierarchy of
quartic couplings 
turn out to be stable
against renormalisation group evolution.
However, taking the
Yukawa couplings into account
$\mu_1>\mu_2$ is
stable but not necessarily $\mu_2>\mu_3$.
Reaching $\mu_2=\mu_3$ at a certain
renormalisation scale would elevate
the CP2 symmetry of the Higgs potential to
a CP3 symmetry. But, of course, this does
{\em not} imply that the full MCPM which
includes fermions and gauge bosons has
a higher symmetry than CP2 at this
renormalisation scale.

Eventually, we have computed the oblique parameters
in the MCPM. We find for large parameter
space agreement with the
electroweak precision measurements. 
In particular we have presented
the 1-$\sigma$ and 2-$\sigma$
contours of valid regions in
the $m_{h'}$--$m_{h''}$ mass plane for
different choices for the charged-Higgs-boson mass and for
SM-like Higgs-boson masses
of 125~GeV and 170~GeV, respectively.
The allowed regions for the masses of
the Higgs bosons are in a reasonable 
range; see Figures~\ref{figob} and
\ref{figob2}. These Higgs bosons 
with masses below 500~GeV should therefore
be detectable in the LHC experiments.
As shown in~\cite{Maniatis:2009vp,Maniatis:2009by,Maniatis:2010sb}
in the MCPM these Higgs bosons have
characteristic production and
decay properties giving
clear experimental signatures.

\appendix
%
\section{Derivation of symmetry classes}
\label{appA}

In this appendix we give a recipe which allows an easy
identification of the symmetry class of any
given THDM potential~\eqref{eqVK}.

The first step is to diagonalise~$E$ by a 
basis transformation~\eqref{eqbasistfK}.
We get then
\begin{equation}
E = \diag (\mu_1, \mu_2, \mu_3 ).
\end{equation}
Since~$E$ is symmetric a diagonalisation is always possible.
Therefore we work in the following in the $E$ diagonal basis
and consider~$\tvec{\xi}$ and
$\tvec{\eta}$~\eqref{connect} in this basis.
Now we have to distinguish three cases for the~$\mu$'s.

\begin{itemize}
\item[(a)] $\mu_1$, $\mu_2$, $\mu_3$ all different.

Then we see from~\eqref{eqR} and \eqref{eqRbar}
that only diagonal O(3) matrices $R$ or $\bar{R}$
may lead to symmetries, that is, we have to consider
\begin{equation}
\label{casesa}
\begin{split}
\bar{R} &= \bar{R}_j \quad (j=1,2,3; \quad \text{see}~\eqref{eqreflect} ),\\
\bar{R} &= \bar{R}_1 \bar{R}_2 \bar{R}_3 = - \unitmatrix_3 ,\\
R	&= \bar{R}_i \bar{R}_j \quad \text{with } i \neq j .
\end{split}
\end{equation}
Now we can easily check the conditions for~$\tvec{\xi}$,
$\tvec{\eta}$ from~\eqref{eqR} and \eqref{eqRbar}. We can 
have the following cases:

\begin{itemize}
\item[(a.1)] $(\xi_i,\eta_i) \neq (0,0) \quad \text{for } i=1,2,3 $.

With none of the matrices from~\eqref{casesa} we can 
fulfill the symmetry relations in~\eqref{eqR}.
In this case the potential has only the trivial symmetry
group~$\mathbbm{G}=\{ \unitmatrix_3 \}$.

\item[(a.2)] Exactly one pair fulfills $(\xi_i, \eta_i) = (0,0)$ 
where $i \in \{1,2,3 \}$.

Without loss of generality we can set $(\xi_2, \eta_2) = (0,0)$,
$(\xi_1, \eta_1) \neq (0,0)$,
$(\xi_3, \eta_3) \neq (0,0)$.
Clearly, from~\eqref{eqR}, \eqref{eqRbar} we have
$\bar{R}_2$ and nothing else as symmetry transformation.
We get the symmetry group 
\begin{equation}
\mathbbm{G} =\{ \unitmatrix_3, \bar{R}_2 \}
\end{equation}
which characterises the CP1 symmetry class; see Table~\ref{tabsymmgroup},
the first subclass of CP1.

\item[(a.3)] Exactly two pairs fulfill $(\xi_i, \eta_i) = (0,0)$ 
where $i \in \{1,2,3 \}$.

Without loss of generality we can set $(\xi_1, \eta_1)=(\xi_2, \eta_2) = (0,0)$,
$(\xi_3, \eta_3) \neq (0,0)$.
From~\eqref{eqR}, \eqref{eqRbar} and~\eqref{casesa} we see that
the invariance group is 
\begin{equation}
\mathbbm{G} =\{ \unitmatrix_3, \bar{R}_1, \bar{R}_2, \bar{R}_1 \bar{R}_2 \} .
\end{equation}
We get the symmetry group characterising the $\mathbbm{Z}_2$ symmetry
class; see Table~\ref{tabsymmgroup}, the first subclass
of $\mathbbm{Z}_2$.

\item[(a.4)] $\tvec{\xi}=0$ and $\tvec{\eta}=0$ .

Here we find from~\eqref{eqR}, \eqref{eqRbar} and~\eqref{casesa} as
symmetry group
\begin{equation}
\mathbbm{G} =\{ \unitmatrix_3, \bar{R}_1, \bar{R}_2, \bar{R}_3, \bar{R}_1 \bar{R}_2
, \bar{R}_2 \bar{R}_3, \bar{R}_1 \bar{R}_3,
\bar{R}_1 \bar{R}_2 \bar{R}_3 = - \unitmatrix_3 
\} .
\end{equation}
This characterises the CP2 symmetry class.
\end{itemize}

\item[(b)] Exactly two eigenvalues $\mu_j$ of $E$ are equal.

Without loss of generality we set 
\begin{equation}
\label{eqmudeg}
\mu_1=\mu_2 \neq \mu_3 .
\end{equation}
From ~\eqref{eqR}, \eqref{eqRbar} we see that~$E$
allows now as invariances
\begin{align}
\label{eqA11}
R_{2\theta} &= 
\begin{pmatrix}
\cos (2\theta) & -\sin (2\theta) & 0 \\
\sin (2\theta) & \cos (2\theta)  & 0 \\
0 & 0 & 1
\end{pmatrix}
\qquad 0 \le \theta < \pi ,\\
\label{eqA12}
\bar{R} &= \bar{R}_j, \quad j=2,3  ,\\
\label{eqA13}
\bar{R} &= R_{2\theta} \bar{R}_3 ,\\
\label{eqA14}
\bar{R} &= R_{2\theta} \bar{R}_2 .
\end{align}
Note that $\bar{R}_1$ is included in~\eqref{eqA14} for $2\theta=\pi$:
$\bar{R}_1 = R_\pi \bar{R}_2$.

Now we consider again all possibilities for~$\tvec{\xi}$ and $\tvec{\eta}$.

\begin{itemize}
\item[(b.1)] $(\xi_3, \eta_3 ) \neq (0,0)$ and
$(\tvec{\xi} \times \tvec{\eta}) \tvec{e}_3 \neq 0$ .

The first relation implies that neither $\bar{R}_3$~\eqref{eqA12} nor any
$R_{2\theta} \bar{R}_3$~\eqref{eqA13} can lead to a
symmetry. The second relation implies  that 
$(\xi_1, \xi_2)^\trans$ and
$(\eta_1, \eta_2)^\trans$ are linearly independent. Therefore,
neither $R_{2\theta}$~\eqref{eqA11} nor any
$R_{2\theta} \bar{R}_2$~\eqref{eqA14}
can lead to a symmetry and
we have here only the trivial invariance group
\begin{equation}
\mathbbm{G} =\{ \unitmatrix_3 \} .
\end{equation}

\item[(b.2)] $(\xi_3, \eta_3) \neq (0,0)$,
$(\tvec{\xi} \times \tvec{\eta}) \tvec{e}_3 = 0$,
$(\tvec{\xi} - \xi_3 \tvec{e}_3, \tvec{\eta} - \eta_3 \tvec{e}_3) \neq (0,0)$.

Here the vectors
$(\xi_1, \xi_2)^\trans$ and
$(\eta_1, \eta_2)^\trans$ are linearly dependent but
at least one of them is non zero. Due to $\mu_1=\mu_2 \neq \mu_3$,
see~\eqref{eqmudeg},
we can make a basis change 
in the 1--2 subspace and achieve, without loss
of generality,
$(\xi_1, \eta_1) \neq (0,0)$ and
$(\xi_2, \eta_2) = (0,0)$.
We see now that here from all possible invariances~\eqref{eqA11} 
to \eqref{eqA14}
only $\bar{R}_2$ remains. Thus, the invariance group is
\begin{equation}
\mathbbm{G} = \{ \unitmatrix_3, \bar{R}_2 \}
\end{equation}
and we get the CP1 class. This is the third subclass of CP1 listed
in Table~\ref{tabsymmgroup}.

\item[(b.3)] $(\xi_3, \eta_3) \neq (0,0)$,
$(\tvec{\xi} \times \tvec{\eta}) \tvec{e}_3 = 0$,
$(\tvec{\xi} - \xi_3 \tvec{e}_3, \tvec{\eta} - \eta_3 \tvec{e}_3) = (0,0)$.

This case can also be characterised by
\begin{equation}
\tvec{\xi} \times \tvec{e}_3 =0 , \quad
\tvec{\eta} \times \tvec{e}_3 =0 , \quad
(\tvec{\xi}, \tvec{\eta}) \neq (0,0) .
\end{equation}
That is, we have here
\begin{equation}
\tvec{\xi} = \begin{pmatrix} 0\\ 0\\ \xi_3 \end{pmatrix}, \quad
\tvec{\eta} = \begin{pmatrix} 0\\ 0\\ \eta_3 \end{pmatrix}, \quad
(\xi_3, \eta_3) \neq (0,0) .
\end{equation}
From~\eqref{eqA11} to \eqref{eqA14} we see that in this case
the invariance group is
\begin{equation}
\mathbbm{G} = \{ R_{2\theta}, R_{2\theta} \bar{R}_2 \} .
\end{equation}
We get the first subclass of the U(1) symmetry class
in Table~\ref{tabsymmgroup}.

\item[(b.4)] $(\xi_3, \eta_3) = (0,0)$,
$\tvec{\xi} \times \tvec{\eta} \neq 0$ .

Here
$(\xi_1, \xi_2)^\trans$ and
$(\eta_1, \eta_2)^\trans$ are linearly independent.
We see from~\eqref{eqR} and \eqref{eqRbar} that neither
$R_{2\theta}$ ($0<\theta<\pi$) nor
$R_{2\theta} \bar{R}_2$ ($0 \le \theta<\pi$) can lead to invariances.
But, clearly, $\bar{R}_3$ gives an invariance and
the corresponding symmetry group is
\begin{equation}
\mathbbm{G} = \{ \unitmatrix_3, \bar{R}_3 \} .
\end{equation}
This group is, of course, equivalent to
$\mathbbm{G} = \{ \unitmatrix_3, \bar{R}_2 \}$ as we see
after a trivial exchange of numbering of the
2 and the 3 axes. We list this case as second
subclass of the CP1 class 
in Table~\ref{tabsymmgroup}.

\item[(b.5)] $(\xi_3, \eta_3) = (0,0)$,
$\tvec{\xi} \times \tvec{\eta} = 0$,
$(\tvec{\xi}, \tvec{\eta}) \neq (0,0)$.

Here
$(\xi_1, \xi_2)^\trans$ and
$(\eta_1, \eta_2)^\trans$ are linearly dependent.
We can make a rotation in the 1--2 subspace to
achieve $(\xi_1, \eta_1) \neq (0,0)$ and
$(\xi_2, \eta_2) = (0,0)$. From~\eqref{eqR},
\eqref{eqRbar} and \eqref{eqA11} to \eqref{eqA14}
we see that here the symmetry group is
\begin{equation}
\mathbbm{G} = \{ \unitmatrix_3, \bar{R}_2, \bar{R}_3, \bar{R}_2 \bar{R}_3 \} .
\end{equation}
After an exchange of numbering of the 1 and 3 axes
this gives the second subclass of the $\mathbbm{Z}_2$ class
in Table~\ref{tabsymmgroup}.

\item[(b.6)] $\tvec{\xi}=0$, $\tvec{\eta}=0$.

Here we have invariance for all the transformations
\eqref{eqA11} to \eqref{eqA14}. The corresponding
symmetry group is
\begin{equation}
\mathbbm{G} = \{ R_{2\theta}, R_{2\theta} \bar{R}_2,
 R_{2\theta} \bar{R}_3 \} \quad \text{with } 0 \le \theta < \pi.
 \end{equation}
We get the CP3 class.

\end{itemize}

\item[(c)] $\mu_1=\mu_2=\mu_3 \equiv \mu$.

Here we have $E = \mu \unitmatrix_3$ and $E$ allows as
invariance all $R$ and $\bar{R}$ matrices of O(3).
We distinguish the following subcases.

\begin{itemize}
\item[(c.1)] $\tvec{\xi} \times \tvec{\eta} \neq 0$.

Without loss of generality we choose the second axis
to be parallel to $\tvec{\xi} \times \tvec{\eta}$.
We have then $(\xi_2, \eta_2) = (0,0)$ and further that
$(\xi_1, \xi_3)^\trans$ and
$(\eta_1, \eta_3)^\trans$ are linearly independent.
The invariance group is
\begin{equation}
\mathbbm{G} = \{ \unitmatrix_3, \bar{R}_2  \}
\end{equation}
and we get the fourth subclass of the CP1 class
in Table~\ref{tabsymmgroup}.

\item[(c.2)] $\tvec{\xi} \times \tvec{\eta} = 0$,
$(\tvec{\xi}, \tvec{\eta}) \neq (0,0)$.

Here the vectors $\tvec{\xi}$ and $\tvec{\eta}$
are parallel and at least one of them is
unequal zero. We choose this vector to 
define the~3~axis and get
\begin{equation}
\tvec{\xi} = \begin{pmatrix} 0\\ 0\\ \xi_3 \end{pmatrix} ,\quad
\tvec{\eta} = \begin{pmatrix} 0\\ 0\\ \eta_3 \end{pmatrix} ,\quad
\text{with } (\xi_3, \eta_3) \neq (0,0) .
\end{equation}
The invariance group is then from~\eqref{eqR}
and \eqref{eqRbar}
\begin{equation}
\mathbbm{G} = \{ R_{2\theta}, R_{2\theta} \bar{R}_2  \};
\end{equation}
see \eqref{eqA11} to \eqref{eqA14}.
We get the second subclass of the U(1) class
in Table~\ref{tabsymmgroup}.

\item[(c.3)] $\tvec{\xi}=0$, $\tvec{\eta}=0$.

Here, clearly, we get as symmetry group
\begin{equation}
\mathbbm{G} = \text{O(3)} = 
\{ R, R \bar{R}_2  \}, \quad \text{with } R \in \text{SO(3)}.
\end{equation}
This is labeled as SO(3) class
in Table~\ref{tabsymmgroup}.

\end{itemize}
\end{itemize}

To summarize, in this appendix we have -- in a systematic
way -- gone through all possibilities for the potential
parameters~$E$, $\tvec{\xi}$, $\tvec{\eta}$ and checked
for possible symmetry groups. For any given THDM
potential all these steps are easily done and this
gives a practical way to identify if any and what
symmetry the potential has.
Of course, a symmetry of the potential is not
guaranteed to be respected by the Yukawa couplings.
This has to be checked as a second step.
Such a program has, for instance, be carried
through for the MCPM in~\cite{Maniatis:2007de}.\\

The corespondence of the Higgs-family transformations for fields
and field bilinears is given in~\eqref{eq11} resp. \eqref{eqbasistfK}.
An~$R$ in \eqref{eqbasistfK} determines $U$ in \eqref{eq11} up to
gauge transformations.
Similarly, for GCP transformations~$\bar{R}$ in~\eqref{eqGCPK} determines
$U$ in \eqref{eqX} up to gauge transformations. 
In Tables~\ref{tabRU} and \ref{tabRbarU}
we give these correspondences of transformations in field
and $K$ space for the elements of the groups~$\mathbbm{G}$ occuring
in Table~\ref{tabsymmgroup}.
\begin{table*}[t!]
\centering
\begin{tabular}{ll}
\hline
$R$ & $U$ \\
\hline
$\bar{R}_1 \bar{R}_2 = \diag (-1,-1,1) $
&
$\sigma^3$ \\
$\bar{R}_2 \bar{R}_3 = \diag (1,-1,-1) $
&
$\sigma^1$ \\
$\bar{R}_1 \bar{R}_3 = \diag (-1,1,-1) $
&
$\sigma^2$ \\
$R_{2\theta}$, see \eqref{eqA11} and \eqref{eqrot}
with $\alpha=2\theta$ \qquad \qquad
&
$\cos (\theta) \unitmatrix_2 - i \sin (\theta) \sigma^3 $\\
\hline
\end{tabular}
\caption{\label{tabRU} \small 
Correspondence of proper rotation matrices~$R$
in \eqref{eqbasistfK} and field
transformations~$U$ in~\eqref{eq11}.}
\end{table*}

\begin{table*}[t!]
\centering
\begin{tabular}{ll}
\hline
$\bar{R}$ & $U$ \\
\hline
$\bar{R}_1 = \diag (-1,1,1) $ \qquad \qquad
&
$\sigma^3$ \\
$\bar{R}_2 = \diag (1,-1,1) $
&
$\unitmatrix_2$ \\
$\bar{R}_3 = \diag (1,1,-1) $
&
$\sigma^1$ \\
$-\unitmatrix_3$
&
$\epsilon = i \sigma^2$\\
$R_{2\theta} \bar{R}_2$, see \eqref{eqA14}
&
$\cos (\theta) \unitmatrix_2 - i \sin (\theta) \sigma^3$\\
$R_{2\theta} \bar{R}_3$, see \eqref{eqA13}
&
$\cos (\theta) \sigma^1 + \sin (\theta) \sigma^2$\\
\hline
\end{tabular}
\caption{\label{tabRbarU} \small 
Correspondence of improper rotation matrices~$\bar{R}$
in \eqref{eqGCPK} and matrices
$U$ in GCP transformations of fields~\eqref{eqX}.}
\end{table*}

%
\section{Lagrangian of the MCPM}
\label{appLag}

In sections~\ref{secren} and \ref{secob} we consider a model
corresponding to the symmetry class CP2 in Table~\ref{tabsymmgroup},
the MCPM. Here we recall the Lagrangian of this model as
originally given in~\cite{Maniatis:2007de}.

The Lagrangian of the MCPM can be written as
\begin{equation}\label{2.2}
\mathscr{L}_{\text{MCPM}} = \mathscr{L}_\varphi + \mathscr{L}_\text{Yuk} + 
  \mathscr{L}_\text{FB}\,.
\end{equation}
Here $\mathscr{L}_\text{FB}$ is the standard gauge kinetic Lagrange density
for fermions and gauge bosons (see for instance~\cite{Nachtmann:1990ta}).

The Higgs-boson Lagrangian is 
\begin{equation}\label{2.3}
\mathscr{L}_\varphi = \sum_{i=1,2}
  \left( D_\mu \varphi_i \right)^\dagger \left( D^\mu \varphi_i \right)
  - V(\varphi_1,\varphi_2)\,,
\end{equation}
with $V(\varphi_1,\varphi_2)$
the Higgs potential~\eqref{eqVK} with the constraints~\eqref{MCPMpot}.
The covariant derivative reads
\begin{equation}
D_\mu = \partial_\mu 
+ i g_2 W_\mu^a {\bf T}_a
+ i g_1 B_\mu {\bf Y}
\end{equation}
where~${\bf T}_a$ and ${\bf Y}$ are the generating operators
of weak-isospin and weak-hypercharge transformations,
respectively. 
$W_\mu^a$, $a=1,2,3$ and $B_\mu$ are the gauge fields
and $g_2$ and $g_1$ the corresponding gauge couplings.
For the Higgs doublets we have ${\bf T}_a = \sigma^a/2$
where $\sigma^a$ with $a=1,2,3$ are the Pauli matrices.
We choose the convention that both Higgs-boson doublets
have weak hypercharge $y=+1/2$. 

Furthermore, $\mathscr{L}_\text{Yuk}$ denotes the Yukawa term
which in the 
MCPM has the form
\begin{equation}
\begin{split}
\label{eqYuk}
\mathscr{L}_{\mathrm{Yuk}}(x) = 
  -c_\tau & \;\Bigg[
    \bar{\tau}_{R}(x)\,\varphi_1^\dagger(x)
    \begin{pmatrix} \nu_{\tau\,L}(x) \\ \tau_{L}(x) \end{pmatrix}
\;
    -\bar{\mu}_{R}(x)\,\varphi_2^\dagger(x)
    \begin{pmatrix} \nu_{\mu\,L}(x) \\ \mu_{L}(x) \end{pmatrix}
    \Bigg]
\\
  +c_t &\;\Bigg[
    \bar{t}_{R}(x)\,\varphi_1^\trans(x)\,\epsilon
    \begin{pmatrix} t_{L}(x) \\ b_{L}(x) \end{pmatrix}
\;
    -\bar{c}_{R}(x)\,\varphi_2^\trans(x)\,\epsilon
    \begin{pmatrix} c_{L}(x) \\ s_{L}(x) \end{pmatrix}
    \Bigg]
\\
  -c_b &\;\Bigg[
    \bar{b}_{R}(x)\,\varphi_1^\dagger(x)
    \begin{pmatrix} t_{L}(x) \\ b_{L}(x) \end{pmatrix}
\;
    -\bar{s}_{R}(x)\,\varphi_2^\dagger(x)
    \begin{pmatrix} c_{L}(x) \\ s_{L}(x) \end{pmatrix}
    \Bigg]
		+ h.c.
\end{split}
\end{equation}
where $\epsilon=i \sigma^2$ and $c_\tau$, $c_t$ and $c_b$ 
are real positive constants, determined by the vacuum
expectation value~$v_0$ and
the fermion masses; see \eqref{tbmasses}.
Note that the first family remains uncoupled 
-- at tree level -- to the Higgs bosons in the MCPM.

Through EWSB
only the Higgs-boson doublet $\varphi_1$ gets a vacuum-expectation value. 
In the unitary gauge we have
\begin{equation}
\label{equnit}
\varphi_1(x) =
\frac{1}{\sqrt{2}}
\begin{pmatrix}
0 \\ v_0 + \rho'(x)
\end{pmatrix},
\qquad
\varphi_2(x) =
\begin{pmatrix}
H^+(x) \\ 
\frac{1}{\sqrt{2}} ( h'(x) + i h''(x) )
\end{pmatrix},
\end{equation}
where $\rho'(x)$, $h'(x)$ and $h''(x)$ are the 
real fields corresponding to the physical neutral Higgs 
particles. The fields $H^+(x)$ and 
$H^-(x) = \big( H^+(x)\big)^\dagger$
correspond to the physical charged Higgs-boson pair.

%
%

\end{document}